\documentclass{ws-procs9x6}

\begin{document}

\title{Structure of the $\sigma$-meson and diamagnetism of the nucleon
\footnote{\uppercase{T}his work is supported by \uppercase{D}eutsche
 \uppercase{F}orschungsgemeinschaft through the projects 
\uppercase{SCHU}222/25 and  436 \uppercase{RUS} 113/510}}

\author{M. I. Levchuk}

\address{B.I. Stepanov Institute of Physics, BY-220072 Minsk, Belarus}

\author{A.I. L'vov}

\address{P.N. Lebedev Physical Institute, RU-117924 Moscow, Russia}  

\author{A.I. Milstein}

\address{Budker Institute of Nuclear Physics, RU-630090 Novosibirsk, Russia}  

\author{M. Schumacher\footnote{\uppercase{C}orresponding author:
e-mail mschuma3@gwdg.de}}

\address{Zweites Physikalisches Institut der Universit\"at, D-37077
  G\"ottingen, Germany}
\maketitle

\abstracts{The structure of the $\sigma$ meson and the 
diamagnetism of the nucleon 
are shown to be  topics which are closely related to  each other. 
Arguments are found that the $\sigma$ meson couples to two photons via its
non-strange $q\bar{q}$ structure component. This ansatz leads 
to a quantitative 
explanation of the $t$-channel component of the difference of electromagnetic
polarizabilities, $(\alpha-\beta)^t$, containing the diamagnetism of the
nucleon.  The prediction is 
$(\alpha-\beta)^t_{p,n}=(5 \alpha_e g_{\pi NN})/(6\pi^2 m^2_\sigma f_\pi)
=15.3$ in units of
$10^{-4}{\rm fm}^3$ to be compared with the experimental values
$(\alpha-\beta)^t_{p}=15.1\pm 1.3$ for the proton and 
$(\alpha-\beta)^t_{n}=14.8\pm 2.7$ for the neutron.
The equivalent approach to exploit the $\pi\pi$ structure component
of the $\sigma$ meson via the BEFT sum rule leads to 
$(\alpha-\beta)^t_{p,n}=14\pm 2$, what also is in agreement with the
experimental results.}

\section{Introduction}

Diamagnetism is one of the dominant properties of the nucleon,
contrasting with the fact that this phenomenon still is barely understood.
A clearcut information  is provided by
dispersion theory\cite{lvov97,drechsel03,wissmann04,schumacher05b} 
where  Compton scattering is described by 
six invariant amplitudes  $A_i(s,t)$. The amplitudes   
$A_i(s,t)$ are analytic functions in the two complex planes $s$ and $t$
and may be calculated from the singularities contained in these planes.
The $s$-plane singularities 
are given by  the meson-photoproduction
cross sections contributing to the total photoabsorption cross section. 
The $t$-channel singularities 
were identified in the late 1950th and
the beginning of the 1960th by Low, Jacob and Mathews (LJM\cite{low58}) 
and Hearn and Leader (HL\cite{hearn62}). These singularities reflect 
the degrees of freedom (d.o.f.)
of the nucleon and we  find it convenient to introduce the terms
$s$-channel d.o.f and $t$-channel d.o.f.
Translated into modern language,
LJM\cite{low58} argued that instead of an excitation
of the pion and constituent-quark structures of the nucleon, i.e. the
$s$-channel d.o.f., the production
of a $\pi^0$ meson in the intermediate state may lead to Compton scattering. 
This $\pi^0$-pole contribution is now
generally accepted as part of the backward spin-polarizability $\gamma_\pi$.
On the other hand the scalar-isoscalar $\pi\pi$ $t$-channel introduced by
HL\cite{hearn62} making a large contribution to
$(\alpha-\beta)$ is  frequently ignored or insufficiently 
represented in theoretical approaches (see Ref.\cite{schumacher05b}).

\section{Experimental status of electromagnetic polarizabilities} 

According to our
recent analysis\cite{schumacher05a,schumacher05b} the experimental 
polarizabilities may be summarized in the
form given in Table \ref{tablepolarizabilities}.
\begin{table}[ph]
\tbl{Summary on electromagnetic polarizabilities 
in units of $10^{-4}{\rm fm}^3$}
{\footnotesize
\begin{tabular}{@{}llll@{}}
\hline
1&&\,\,\quad\quad\quad  proton& \,\,\quad\quad\quad  neutron\\
\hline
2&BL sum rule & $(\alpha+\beta)_p=13.9\pm 0.3$&$
(\alpha+\beta)_n=15.2\pm 0.5$\\
3&Compton scattering&$(\alpha-\beta)_p=10.1\pm 0.9$&
$(\alpha-\beta)_n=9.8\pm 2.5$\\
4&BEFT sum rule&$(\alpha-\beta)^s_p=-5.0\pm 1.0$&$(\alpha-\beta)^s_n=-5.0\pm
1.0$\\ 
5&line 2 -- line 3&
 $(\alpha-\beta)^t_p=15.1\pm 1.3$& $(\alpha-\beta)^t_n=14.8\pm
2.7$\\ 
\hline
6&experimental& $\alpha_p=12.0\pm 0.6$ 
&  $\alpha_n=12.5  \pm 1.7$ \\
7&$s$-channel only & $\alpha^s_p= \,\,\,4.5\pm 0.5$ 
&  $ 
\alpha^s_n=\,\,\,5.1\pm 0.6$\\
8&$t$-channel only &   $\alpha^t_p= \,\,\,7.5\pm 0.8$ 
&  $ 
\alpha^t_n=\,\,\,7.4\pm 1.8$\\
\hline
9&experimental& $\beta_p=\,\,\,1.9\mp 0.6$ 
& $\beta_n=\,\,2.7\mp 1.8$ \\
10&$s$-channel only& $\beta^s_p= \,\,\,9.5\pm 0.5$ & 
 $ 
\beta^s_n=10.1\pm 0.6$\\
11&$t$-channel only& $\beta^t_p= \,\,\,-7.6\pm 0.8$ & 
 $ 
\beta^t_n=-7.4\pm 1.9$\\
\hline
\end{tabular}
\label{tablepolarizabilities}}
\end{table}
The quantities
$\alpha_p,\beta_p,\alpha_n,\beta_n$ are the experimental electric and magnetic
polarizabilities  for the proton and neutron, respectively. The 
quantities with an upper label $s$ are the corresponding electric and magnetic
polarizabilities where only the $s$-channel d.o.f. are included. 
These latter quantities have been obtained by making use
of the fact that $(\alpha+\beta)$, when calculated from forward-angle
dispersion theory as given by the Baldin or Baldin-Lapidus (BL)
sum rule
\begin{equation}
(\alpha+\beta)=\frac{1}{2\pi^2}\int^\infty_{\omega_0}
\frac{\sigma_{\rm tot}(\omega)}{\omega^2}d\omega,
\label{BLsumrule}
\end{equation}
has no $t$-channel contribution, i.e. $(\alpha+\beta)=(\alpha+\beta)^s$,  
and by using
the estimate $(\alpha-\beta)^s_{p,n}=-5.0\pm 1.0$ 
obtained form the $s$-channel part of the BEFT sum rule 
\begin{equation}
(\alpha-\beta)^s=\frac{1}{2\pi^2}\int^\infty_{m_\pi+\frac{m^2\pi}{2M}}
\sqrt{1+\frac{2\omega}{M}}\left[\sigma(\omega,E1,\cdots)
-\sigma(\omega,M1,\cdots)\right]\frac{d\omega}{\omega^2}
\label{BEFTsumrule}
\end{equation}
both for the proton and the
neutron (see Ref.\cite{schumacher05b}), where $M$ is the nucleon mass.  
The numbers in  line 5
of Table \ref{tablepolarizabilities} are the $t$-channel contributions to 
$(\alpha-\beta)$ obtained from the experimental values 
$(\alpha-\beta)_p=10.1\pm 0.9$ and $(\alpha-\beta)_n=9.8\pm 2.5$ and the
estimate for    $(\alpha-\beta)^s_{p,n}$. We see that the 
experimental values for
$\alpha$ are much larger than the $s$-channel contributions alone, whereas for
the magnetic polarizabilities the opposite is true. For the magnetic
polarizability it makes sense to identify the large difference between
the experimental value and the $s$-channel contribution with the diamagnetic
polarizability.    This difference is filled up  by  $\beta^t$                
which, therefore, may be considered as the diamagnetic 
polarizability. It is important to notice that this definition
of the diamagnetic polarizability has nothing in common with the 
``classical'' diamagnetic term given e.g. in  Eq. (12) of 
Ref.\cite{schumacher05b}. This latter term makes use of $s$-channel
degrees of freedom only and, therefore, does  not describe the
physical origin of the diamagnetism.

\section{The $\sigma$-pole and the BEFT sum rule}

The $\sigma$ meson having  quantum numbers $I=0$ and $J=0$
may be considered as a quasi-stable 
$1/\sqrt{2}(|u\bar{u}\rangle+|d\bar{d}\rangle$)
      $1^3{\rm P}_0$ state in a confining 
potential which is coupled to  a di-pion state in the continuum
$1/\sqrt{3}(|\pi^+\pi^-\rangle-|\pi^0\pi^0\rangle+|\pi^-\pi^+\rangle)$,       
showing up as a relative  $S-$wave of the two pions. By exploiting the
non-strange $q\bar{q}$ structure component we  are led  to the 
following relations for the $\sigma$-pole
\begin{eqnarray}
&F_{\sigma\gamma\gamma}&=|M(\sigma\to2\gamma)|=
\frac{\alpha_e}{\pi f_\pi}N_c \left[
\left(\frac23\right)^2 +  \left(-\frac13\right)^2  \right]=\frac53
\frac{\alpha_e}{\pi f_\pi}\label{width3}\\
&(\alpha-\beta)^t_{p,n}&=\frac{g_{\sigma NN}F_{\sigma\gamma\gamma}}{2\pi 
m^2_\sigma}
=\frac{5\alpha_e g_{\pi NN}}{6\pi^2 m^2_\sigma f_\pi}
=15.3\times 10^{-4}{\rm fm}^3
\end{eqnarray}
where $\alpha_e=1/137.04$, $N_c=3$, $m_\sigma=665$ MeV,
$f_\pi=(92.42\pm 0.26)$ MeV and 
$g_{\sigma NN}= g_{\pi NN}= 13.169 \pm 0.057$ Ref.\cite{nagy04}.
The $\pi\pi$ structure component leads to the 
$t$-channel part of the BEFT sum rule in the following form
\begin{equation}
\begin{split}
(\alpha-\beta)^t&= \frac{1}{16\pi^2}\int^\infty_{4m^2_\pi}
\frac{dt}{t^2}\frac{16}{4M^2-t}\left(\frac{t-4m^2_\pi}{t}\right)^{1/2}\\
&\times\Big[ f^0_+(t)F^{0*}_0(t)
-\left(M^2-\frac{t}{4}\right)\left(\frac{t}{4}-m^2_\pi\right)
f^2_+(t)F^{2*}_0(t)
\Big]
\end{split}
\end{equation}
\noindent
where $f^{J=0,2}_+(t)$ is the partial wave amplitude 
of the process $N\bar{N}\to\pi\pi$ and 
$F^{J=0,2}_{I=0}(t)$ the partial wave amplitude of the process
$\pi\pi\to\gamma\gamma$. The predictions of the BEFT sum rule are
$(\alpha-\beta)^t_{p,n}=+(14\pm2)$ (Levchuk et al., 
see Ref.\cite{schumacher05b}),
$(\alpha-\beta)^t_p=+16.5$ (Drechsel et al.\cite{drechsel03}) with the 
arithmetic average $(\alpha-\beta)^t_{p,n}= +15.3\pm 1.3$.\\
\begin{table}[ph]
\tbl{Difference of  
electromagnetic polarizabilities $(\alpha-\beta)^t_{p,n}$
in units of $10^{-4}{\rm fm}^3$}
{\footnotesize
\begin{tabular}{@{}lll@{}}
\hline
&$(\alpha-\beta)^t_p$ &   $ (\alpha-\beta)^t_n$         \\
\hline
experiment&$15.1 \pm 1.3$& $14.8 \pm 2.7$\\
$\sigma$-pole& $15.3$&   $15.3$                    \\
BEFT sum rule& $15.3\pm 1.3$&  $15.3\pm 1.3$         \\
\hline
\end{tabular}}
\label{summary}
\end{table}
Table \ref{summary} summarizes the results obtained for 
$(\alpha-\beta)^t_{p,n}$.
It is remarkable that the experimental values are in agreement
with the prediction of the $\sigma$-pole as well as the prediction
of the BEFT sum rule. This leads to the tentative conclusion that both
approaches are equivalent and to a confirmation of the  expression given in
(\ref{width3}) for $F_{\sigma\gamma\gamma}$, treating
the two-photon coupling of the $\sigma$-meson 
analogous to the $\pi^0$-meson case.

\end{document}